\documentclass[%
 reprint,
 amsmath,amssymb,
 aps,nofootinbib,
]{revtex4-1}
\usepackage{bm}
\PassOptionsToPackage{linktocpage}{hyperref}
\usepackage[hyperindex,breaklinks]{hyperref}
\usepackage{enumitem}
\usepackage{slashed}

\renewcommand{\theta}{\vartheta}

\renewcommand{\vec}[1]{\ensuremath{\boldsymbol{#1}}}

\newcommand{\bra}[1]{\ensuremath{\left< #1\,\right|}}
\newcommand{\ket}[1]{\ensuremath{\left|\, #1\right>}}

\usepackage{array}
\usepackage{mathtools}

\usepackage{etoolbox}

\begin{document} 

\title{Critically excited states with enhanced memory and pattern recognition capacities in quantum brain networks: Lesson from black holes.}

\author{Gia Dvali} 
\affiliation{%
Arnold Sommerfeld Center, Ludwig-Maximilians-Universit\"at, Theresienstra{\ss}e 37, 80333 M\"unchen, Germany, 
}%
 \affiliation{%
Max-Planck-Institut f\"ur Physik, F\"ohringer Ring 6, 80805 M\"unchen, Germany
}%
 \affiliation{%
Center for Cosmology and Particle Physics, Department of Physics, New York University, 726 Broadway, New York, NY 10003, USA
}%

\date{\today}

\begin{abstract}
 We implement a mechanism - originally proposed as a model for  
the large memory storage capacity of black holes - in quantum neural networks and show that an exponentially increased capacity of pattern storage and recognition  
 is achieved in certain critically excited states, without involvement of synaptic plasticity.  We consider a simple network of $N$ interconnected quantum neurons with weak excitatory synaptic connections. We show
that for frozen synaptic weights
 there exist the critical states of enhanced memory storage capacity.  These states are achieved thanks to the high excitation levels of some of the neurons, 
which  - despite of feeble synaptic connections  - dramatically lower the 
response threshold of the remaining weaker-excited neurons.   
 As a results, the latter neurons acquire a capacity to store an exponentially large number of patterns within a narrow energy gap. The stored patterns can be recognized and retrieved with perfect response under the influence of arbitrarily soft input stimuli.  In sharp contrast, under the same stimuli the recall  is absent in the ground-state of the system.  The lesson is that 
 the state with the highest micro-state entropy and memory storage capacity is not necessarily 
 a local minimum of energy, but rather an excited  critical state.    
 The considered phenomenon has a smooth classical limit and 
 can serve for achieving an enhanced memory storage capacity  in classical brain networks.

\end{abstract}

\maketitle

 \section{Transfer of ideas from black holes to brains} 
  
  Black holes and human brains are the two creations of nature 
 that are extremely efficient storers of information. 
 It is a legitimate question to ask whether these two seemingly remote systems share some fundamental mechanism for increased  capacity 
 of information storage.  \\ 
 
 Here, we shall ask the following set of questions: 

\begin{itemize}
  \item Is there a counterpart of the black hole mechanism of information storage in 
macroscopic quantum brains?  

  \item  Can a similar mechanism be used for the enhancement of {\it classical}
information storage capacity in classical brains? 

  \item Can it be used for storing {\it quantum data}  in classical brains?  
  
\end{itemize} 

The term {\it classical}  will be given a very precise meaning below. \\
 
 
 It has been hypothesized \cite{DG} that 
 the mechanism behind the black hole information storage is based on the fact that  some of the black hole's microscopic constituents are in a highly-excited (highly occupied) critical state.  The high level of excitation of some constituents, due to their interactions with others, puts these others in a 
 state of a sharply increased capacity of memory storage.  
This increase of memory capacity was proposed in \cite{DG}  to be the underlying mechanism for Bekenstein entropy \cite{Bek} carried by a black hole.   
This phenomenon was referred to as the {\it maximal packing}. 
In other words, the lesson is that the state with largest memory capacity is not necessarily a 
ground-state of the system, but rather a highly excited critical state.  \\

 Moreover, the above mechanism of the enhanced information storage capacity was shown to be operative in ordinary quantum systems,  with bosonic qudits
with attractive (excitatory) connections \cite{DG, goldstone, mischa1}.
  Such a system,  effectively represents 
a quantum ``brain"  of sharply enhanced memory capacity, in which patterns can be encoded and retrieved at an arbitrarily small energy cost.\\

The purpose 
of the present paper is to implement this idea in the framework of artificial quantum neural networks \cite{kak}
and to show that  a very similar mechanism leads to  an exponential increase 
of memory storage and pattern recognition capacities. \\ 

 Our main focus is the energy cost efficiency of the pattern storage and recognition.  That is, we are interested in the states in which the network can store, redial    
and retrieve the maximal variety of patterns at a lowest possible energy cost. 
In the same time, it should be responsive - with a maximal precision - to the input patterns fed via maximally soft external stimuli. 
 Here, the term {\it maximally soft } applies to a stimulus  
 that carries a lowest possible energy.  \\

 \section{Criticality}  
  
   A frequent approach in studying neural networks is to search for 
  the states with most efficient memory storage in form of local minima of 
  a certain energy function.  
  When such minima are absent in an initial network, they are formed  by adjusting synaptic connections in the learning process, via synaptic plasticity.  A well-known example of the efficiency of this method is provided by 
the Hopfield  model \cite{Hopfield}. \\
 
  However, the black holes may be teaching us an alternative lesson \cite{DG}.   Translated in neural network language, the idea is 
 that  the states of enhanced memory capacity are not necessarily the local minima of the energy, but rather the critical extremal points in which some of the neurons are in highly excited states. 
 Under the influence of the excited neurons, the remaining 
neurons - that are with the former neurons in excitatory synaptic connections - acquire a flexibility to 
 store and recall an unusually large number of patterns.  The 
 effect is more sound in the networks in which the strength of the excitatory 
synaptic connections among the neurons is weak, whereas  the excitation levels are high.  In such a case the influence of the excited neurons can lower the excitation threshold energy for other neurons to almost
zero.  This effect creates a 
large number of states available for pattern storage, for an extremely low 
energy cost. \\

 One essential point for this phenomenon is that the neurons should not be restricted to being qubits (i.e., the two-state systems), but rather qudits - the degrees of freedom 
 with large or unlimited number of states. This is necessary in order 
 to allow for their high excitations levels. \\
 
 Several studies\cite{DG,goldstone,mischa1} have shown that this mechanism is operative in many-body systems, such as the attractive bosonic particles placed in a box.  The critical states of sharply increased memory capacity 
 were reached when some of the bosonic modes were in highly excited 
 (highly occupied) states, with their excitation level being inversely proportional to the strength of the attractive connections between the excited and the 
 ``dormant"  modes.
 The latter modes then would acquire an ability to store a large number of patterns in their states.
 \\ 
 
  Our point is that such systems are isomorphic to certain types of quantum neural networks, with different momentum modes playing the role of different neurons, 
their occupation numbers playing the role of the excitation levels  of the 
corresponding neurons  and their interactions playing the role of the synaptic connections. \\
  
   The goal of the present paper is to reduce the phenomenon to its bare essentials and to study it on an example of a simple quantum neural network, without any reference to either many body or black hole physics.  \\

    We consider a simple model of $N$ interconnected quantum 
neurons with weak excitatory synaptic connections. 
We freeze the synaptic weights and treat them as time-independent 
fixed parameters. 
 We show that without the need of synaptic plasticity, the system allows for states with sharply increased memory storage capacity.  
These states are reached when a subset of neurons is highly excited to a certain critical level determined by the inverse strength of fixed synaptic connections. 
This allows other neurons to freely store an exponentially large number 
of patterns  within an arbitrarily narrow energy gap.

  Moreover, the patterns can be retrieved under the influence of 
  arbitrarily weak input stimuli. The retrieved patterns are in one
  to one correspondence with external input patterns.   
   Thus, in this critical state the system can recognize with perfect response  
  an exponentially large number of extremely weak external patterns. 
 In the same time, when put in an unexcited state, the very same network cannot recognize any of the same input patterns. \\

  \section{Network} 
  
   We consider a quantum neural network formed by $N$ neurons, which we shall label by a latin index,  $j=1,2,...N$. 
   The excitation level in a given state will be measured by a quantity $Y_j$, which will denote an expectation value of a corresponding quantum
   operator $\hat{Y}_j$ in that state.  For definiteness, the latter will describe an occupation number 
   operator and can be represented by the standard creation annihilation 
  operators, $\hat{Y}_j = \hat{a_j}^{\dagger}\hat{a_j}$, satisfying the usual commutation relations,
         \begin{equation} 
    [\hat{a_j},\hat{a_k}^{\dagger}] = \delta_{jk}\,, \, \, 
  [\hat{a_j},\hat{a_k}]  =   [\hat{a_j}^{\dagger},\hat{a_k}^{\dagger}] =0\,,   
    \label{algebra} 
 \end{equation}  
 where $\delta_{jk} = \begin{cases}
 0   & \text{for},\,  k \neq j  \\  
  1     & \text{for} \, k=j
\end{cases} \, \, $ is Kronecker delta. 

The space of the states of the system is an infinite dimensional  Fock space spanned over the number eigenstates that are represented by 
Dirac's ket vectors $\ket{y_1,y_2,...y_N}$ that are labeled by 
the eigenvalues of $\hat{Y}_j$-modes:
 $\hat{Y}_j \ket{y_1,y_2,...y_N} = y_j\ket{y_1,y_2,...y_N}$. 
 Obviously, being number-eigenvalues, $y_j$ are non-negative integers.  
 Note the distinction of notations between the eigenvalues, $y_j$, and 
 the expectation values, $Y_j$, of the number operators $\hat{Y}_j$.  
 Of course, when the system is in a given number-eigenstate, the two are equal. 
In any generic quantum state described by a ket vector $\ket{Y}$, it is easy to switch to standard classical neural network characteristics, 
by introducing a state vector 
 \begin{equation} 
 \vec{Y} \, = \, 
   \begin{pmatrix}
   Y_1 \\
    Y_2 \\
    ...\\   
    Y_N  
\end{pmatrix} \,,  
\end{equation} 
where each component is an expectation value
$Y_j \equiv \bra{Y}\hat{Y}_j \ket{Y}$. 

The Hamiltonian of the system is: 
 \begin{equation} 
    \hat{H}  =   \sum_{j=1}^{N} \epsilon_j \hat{Y}_j \, - \, 
   \sum_{j, k =1}^{N}  W_{jk} \hat{Y}_j\hat{Y}_k  \, ,   
 \label{H} 
 \end{equation}  
   where $\epsilon_j $ is the minimal energy gap (energy threshold)  required for exciting   
  neuron $\hat{Y}_j$.  In particular, if the energy transmitted 
  to $\hat{Y}_j$ by an external stimulus 
  is much less than $\epsilon_j $, no response takes place. 
  We shall see this explicitly below. 
  $W_{jk}$  is a synaptic weight  matrix. Since we do not want to consider plasticity,  these shall be treated as fixed, non-dynamical parameters. \\
  
  Since $\hat{Y}_j$-s commute with each other (\ref{algebra}), the matrix $W_{jk}$ is 
  symmetric and real.  We are interested in situation when its elements 
   are semi-positive definite, $W_{jk} \geqslant 0$. With the overall
   minus sign in front of the second term in (\ref{H}), this choice ensures that the connections are 
   {\it excitatory
   in energetic sense}. That is, for each $j,k$ pair of excited neurons 
   the connection terms  $W_{jk}$ lowers the energy cost  
   of the network as  compared to what it would be 
  for  $W_{jk}=0$.
    With such connections, the excitation of any given neuron $Y_j$, effectively lowers the energy thresholds     
 for excitations for all the other neurons $Y_{k\neq j}$ connected to the neuron $Y_j$.  \\  
 
  With  $W_{jk}$ positive, one may worry that the Hamiltinian (\ref{H}) will not 
  be bounded from below, if arbitrary high values of $y_j$-s are allowed.  
   This is not an issue.  It is easy to restrict the maximal values of 
   $y_j$ and/or include higher order stabilizing terms in the Hamiltonian, 
   without affecting any of our conclusions.   The phenomenon we are after  takes place for the states in which such higher order terms 
   can be arbitrarily weak and play no role. \\ 
   
   [In (\ref{H}), we chose a simple number-conserving connection matrix for maximal clarity. Inclusion of number non-conserving connections 
   (e.g., $\hat{a}_j^{\dagger},\hat{a}_j^{\dagger},\hat{a}_k \hat{a}_k$, ...) 
 do not change the essence of the phenomenon.] \\

   The above model is reminiscent of a quantum version of the Hopfield  model \cite{Hopfield}.  However, it is important to understand the crucial differences.   In standard Hopfield model, the neurons would be qubits,  (i.e., two-state systems) and one  would look for local minimal by training the network using 
the synaptic plasticity (see, 
\cite{qHopfield}, and references therein,  for recent analysis in this direction).   \\

 Our idea is different.  We are not interested in adjusting synaptic weights  in the learning process, but rather in understanding of how the system makes the best of it with the given synaptic matrix.  Therefore, first, we take non-plastic synaptic weights. Secondly, we are not looking for the local minima, but rather 
 for critical states with some highly excited neurons.  Finally, for achieving this, it is crucial that our neurons are qudits, i.e., can be in highly excited states. \\  
   
    In order to illustrate our point we shall take the threshold energy gaps to be universal, 
 $\epsilon_j = 1$, which will serve as our unit of energy.  We shall take the synaptic weights to be universal as well, 
 \begin{equation}
 W_{jk} = 
 \begin{cases}
 {g\over 2}    & \text{for},\,  k \neq j  \\  
  0     & \text{for}, \, k = j \,,
\end{cases} \, \, 
 \label{anz} 
 \end{equation}
 where $g$ is a positive parameter. The sign is chosen in order for the connection 
 to be excitatory in energetic sense. We are most interested in the situation in which 
 $g$ is very small and is given by some inverse power of $N$. \\ 

\begin{figure}
 	\begin{center}
        \includegraphics[width=0.6\textwidth]{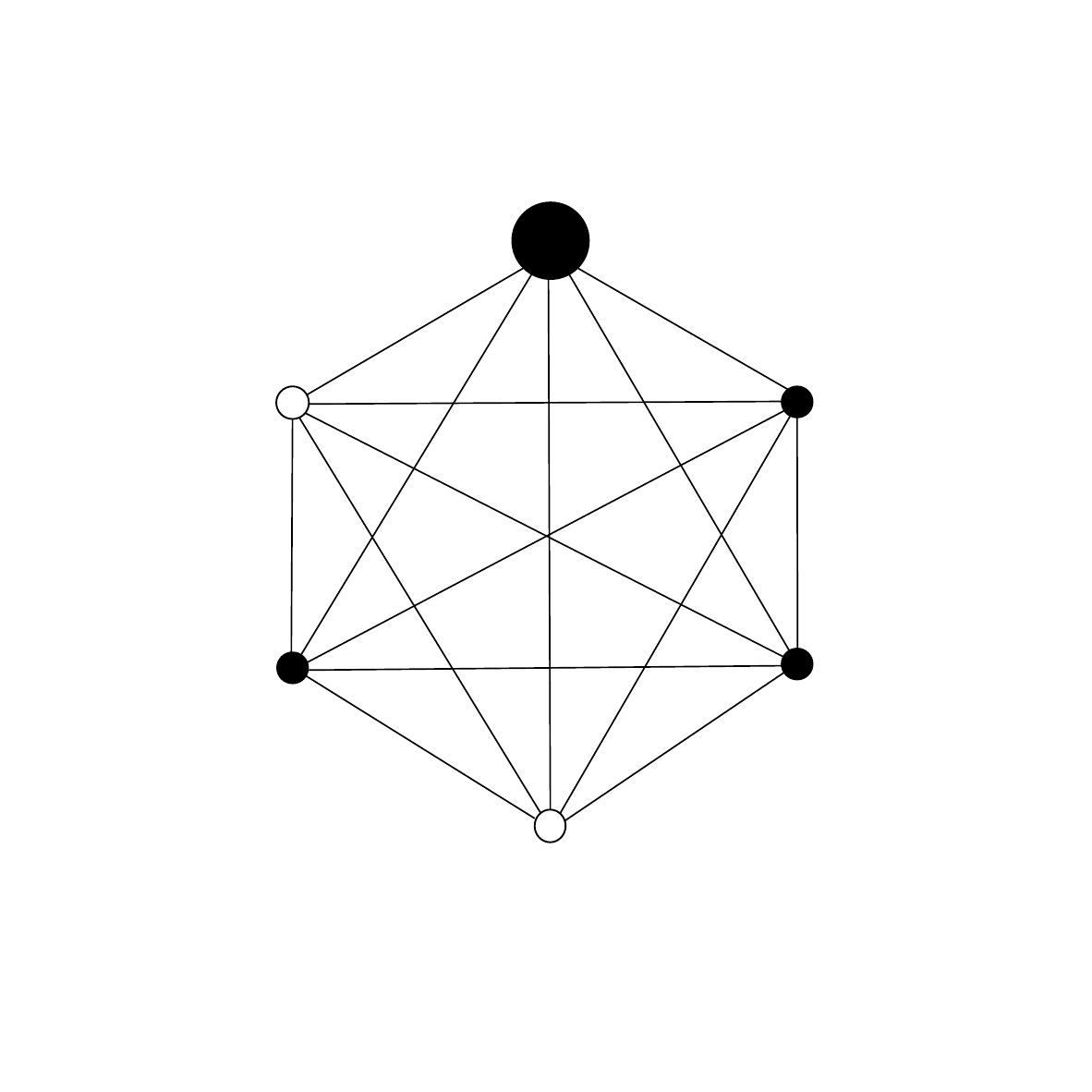}
 		\caption{The neural network representation of a generic state  
of the system described by Hamiltonian (\ref{H1}), for $N=6$.  The black circles represent $Y_j$ neurons, with filled and empty circles denoting excited and unexcited neurons.  The size of a filled circle measures the  excitation
	level of a given neuron. The connecting lines describe synaptic connections
between the different neurons.} 
	\label{Ynet}
 	\end{center}
 \end{figure}

 Thus, the Hamiltonian takes the form: 
   \begin{equation} 
    \hat{H}  =   \sum_{j=1}^{N}  \hat{Y}_j \, - {g \over 2}  \, 
   \sum_{j\neq k}^{N}  \hat{Y}_j\hat{Y}_k  \, .    
 \label{H1} 
 \end{equation}  
 The neural network representation of this system is described by 
 Fig.\ref{Ynet}. \\
 
 We shall now look for the states with maximal memory capacity. 
 The latter we shall quantify as the density of patterns, i.e.,  
   number of patterns that one can store within an unit energy gap. 
Another important characteristics that we are interested in, is the response 
precision with respect to soft external stimuli, i.e., the precision   
by which the network can reproduce different patterns encoded 
in input stimuli of the lowest possible energy.  This is an important measure 
of the energy cost efficiency, both, for  pattern recognition as well as for  information encoding. \\   
   
    If one naively looks in the neighbourhood of the Fock vacuum, 
  $\ket{0, 0, ....0}$ the memory capacity is not especially good. 
  In particular, within the energy gap $\Delta E < 1$, we cannot store any pattern.  Nor, the system can respond to any external pattern encoded in a stimulus of energy $\ll 1$.  Naively, since $g$ is very small, the excitatory synaptic connections cannot change this situation dramatically. 
  However, we are looking in the wrong place. \\
  
  The point is that even extremely weak synaptic connections 
 can become very important when some of the neurons are in highly 
 excited states.  Intuitively it is clear that, since $g$ is small, the excitation 
 level must be around $1/g$, in order to achieve a qualitative effect. 
   In order to see this, let us consider the states in which one of the 
   neurons is highly excited, but below $1/g$. 
   So, let us  take a state with $ 0 < y_1 < 1/g$ and the rest unexcited: 
    $\ket{y_1, 0,0, ....0}$.  In such states, the threshold energies required for 
    exciting other neurons are  effectively lowered to   \\
 \begin{equation} 
   \epsilon_{eff} = 1 -  gy_1 \,.  
 \label{LOW}
 \end{equation}  \\
 For example, this is the threshold energy  required for  making 
 a transition $\ket{y_1, 0,0, ....0} \rightarrow  \ket{y_1, 0,1, ....0}$.
   Thus, closer is  $y_1$ to $1/g$, smaller is the energy threshold for exciting 
   $y_{j\neq 1}$ neurons.  The critical state is achieved at 
   $y_1 = {1 \over g}$, when the threshold becomes zero.  At this 
   point  the rest of neurons become effectively {\it gapless}. 
    One can say that they are ``liberated" and can store very large number 
    of patterns at essentially zero energy cost.  \\

   Let us have a closer look at such critical states. 
  We fix the excitation level of the  first neuron to $y_1 = {1 \over g}$
  and allow some ``excursions" in others. 
   That is, we focus our attention at the states of the form 
  $\ket{\vec{Y}} = \ket{{1 \over g}, y_2, ....y_N}$, where $y_2,...y_N$ can 
  assume variety of values.  \\

  On such states, we can replace 
  $\hat{Y_1}$ by its expectation value  $Y_1=y_1 = {1\over g}$ and the effective Hamiltonian becomes, 
    \begin{equation} 
    \hat{H}  =   {1 \over g}  - {g \over 2}  \, 
   \sum_{1\neq j\neq k \neq 1}^{N}  \hat{Y}_j\hat{Y}_k  \, .    
 \label{H} 
 \end{equation}

  The system governed by the above hamiltonian can store
  an arbitrarily large number of patterns within an unit energy gap,
  provided $g$ is small enough.  The number of patterns in general is exponentially large. The precise number depends on the value of $g$ 
  and the restriction on the maximal values of $y_{j\neq 1}$-neurons 
  one wishes to impose for storing patterns. \\
  
   For a quantitative estimate, we can choose as a reference point
   the state  $\ket{\vec{\xi}} = \ket{{1 \over g}, 0, ....0}$, which is the energy eigenstate with the energy 
   $E_{\vec{\xi}} = {1 \over g}$. The energy difference between it and a state 
   $\ket{\vec{Y}} = \ket{{1 \over g}, y_2, ....y_N}$ is 
    \begin{equation} 
    \Delta E =   |E_{\vec{\xi}} - E_{\vec{Y}} | = 
  {g \over 2} \sum_{1\neq j\neq k \neq 1}^{N} y_jy_k  \,.    
 \label{EEY} 
 \end{equation} 
 In particular, taking combinatorics into account, the energy gap between the states $\ket{\vec{\xi}} = \ket{{1 \over g}, 0, ....0}$ and $\ket{\vec{Y}} = \ket{{1 \over g}, d,d ....d}$, where 
 $d$ is some number, is 
    \begin{equation} 
    \Delta E = 
  g \left ({1 \over 2} \, d^2 (N-1)(N-2) \right ) \,.    
 \label{EEYdd} 
 \end{equation} \\
 
The variety of states $\ket{{1 \over g}, y_2, ....y_N}$, in which 
$y_{j\neq 1}$-s  are arranged in all possible sequences of numbers within 
$0\leqslant  y_{j\neq 1} \leqslant d$, form a library of basic patterns that occupy the above energy gap.  
Their number is, \\
 \begin{equation}
 {\mathcal N}_{pattern} = (d+1)^{N-1} \, . 
 \label{NPatternD}
 \end{equation} \\
Thus, we see that by taking  $g$ sufficiently small and $N$ and $d$ sufficiently large, we can store an {\it arbitrarily large} number of patterns 
within an {\it arbitrarily small}  energy gap. \\

 For example, taking $g = {1 \over N^2}$, the number of 
 patterns with $y_{j\neq1} \leqslant 1$  stored within the unit energy gap 
 is ${\mathcal N}_{pattern} > 2^{N-1}$.  
 Another example:  For   $g = {1 \over N^4}$, the number of patterns 
 with $y_{j\neq1}\leqslant N$,  stored within the same unit energy gap, is
 \begin{equation}
 {\mathcal N}_{pattern} > N^{N-1} \,.
 \label{NPAT}
 \end{equation}
  \\
 
  Such a dramatic increase of memory storage capacity is a trade-off. 
  By putting itself into a highly excited state for some neurons, the system manages to 
  create a huge storage-room for memorizing a large number of patterns within a small energy gap using the rest of  neurons.  \\
  
  The retrieval of patterns from that room is equally efficient.  
  In fact,  patterns can be encoded or retrieved with a perfect response to arbitrarily soft external stimuli. 
  
  \section{Response to input patterns} 
  
   In order to see this, let us put the system under an influence of 
   external stimuli transmitted by the degrees of freedom that we denote by $\hat{X}_j = \hat{b_j}^{\dagger}\hat{b_j}$.
   Again,   $\hat{b_j}^{\dagger}, \hat{b_j}$ are creation/annihilation operators  satisfying  the usual commutation relations, 
           \begin{equation} 
    [\hat{b_j},\hat{b_k}^{\dagger}] = \delta_{jk}\,, \, \, 
  [\hat{b_j},\hat{b_k}]  =   [\hat{b_j}^{\dagger},\hat{b_k}^{\dagger}] =0\,,    
    \label{algebra} 
 \end{equation}  
 and they all commute with analogous $\hat{a_j}^{\dagger},\hat{a_j}$-operators from the $Y$-sector. 
 The $\hat{X}_j$ can be viewed as the neuron of an input layer that is connected to a corresponding output layer neuron $\hat{Y}_j$. 
  After taking into account the interactions, the patterns described by 
  the input layer neurons will become encoded in (or recognized by) the 
  output layer ones. For definiteness, we shall consider the encoding 
  mechanism discussed in \cite{mischa1}.      
 \\
 
  The Hamiltonian now takes the form  
    \begin{equation} 
    \hat{H}  =   \sum_{j=1}^{N}  \hat{Y}_j \, - {g \over 2} \, 
   \sum_{j\neq k}^{N}  \hat{Y}_j\hat{Y}_k  \,  + 
    q \sum_{j=1}^{N}  (\hat{b_j}^{\dagger} \hat{a_j}  + \hat{a_j}^{\dagger} \hat{b_j} )  +  \epsilon \sum_{j=1}^{N}  \hat{X}_j \,,
 \label{HX} 
 \end{equation}  
 where $\epsilon$ measures the energy gap for the minimal excitation level of each input neuron, and $q$ is the strength of the synaptic connection between the input and the output neurons.
 Since we wish the input stimulus to be very soft, we take 
 both $\epsilon$ and $q$ to be very small. In fact, in order to demonstrate the 
 effect for the case of extreme softness, we take $\epsilon =0$.  
 \\
  
  Since $y_1$ is fixed, we shall feed the external stimulus only to 
  the neurons $Y_{j \neq 1}$. 
   For definiteness, we choose that initially 
  $y_{j\neq1} = 0$. That is, as an initial state we choose
  $\ket{\vec{Y}_{in}} = \ket{{1\over g},0,0, ...0}$. 
  We choose the state of external stimulus to be  a coherent state $\ket{\vec{X}_{in}}$, i.e., the eigenvector for all  $\hat{b}_{j}$ 
  annihilation operators, $\hat{b}_{j} \ket{\vec{X}_{in}} = \sqrt{X_j} \ket{\vec{X}_{in}} $, where $\sqrt{X_j}$ is a corresponding eigenvalue. 
   In order not to disturb the value $y_1= {1 \over g}$ we choose $X_1$ to be zero. \\
  
  Note, the choice of a coherent state as opposed to a number eigenstate 
  for the input layer neuron is simply dictated by the fact that such input 
  is easy to generalize for describing an external classical stimulus
  (see below).  \\
  
  Thus, the initial input pattern is given by a vector,
   \begin{equation} 
 \vec{X}_{in} \, = \, 
   \begin{pmatrix}
   0 \\
    X_2 \\
    ...\\   
    X_N  
\end{pmatrix} \,.  
\end{equation} 
We wish to see how the output layer neurons respond to this input pattern. 
 We shall say that the response is perfect if after some time 
 the state vector $\vec{Y}$  can closely repeat the form of the input vector 
  $\vec{X}_{in}$.  \\

\begin{figure}
 	\begin{center}
        \includegraphics[width=0.6\textwidth]{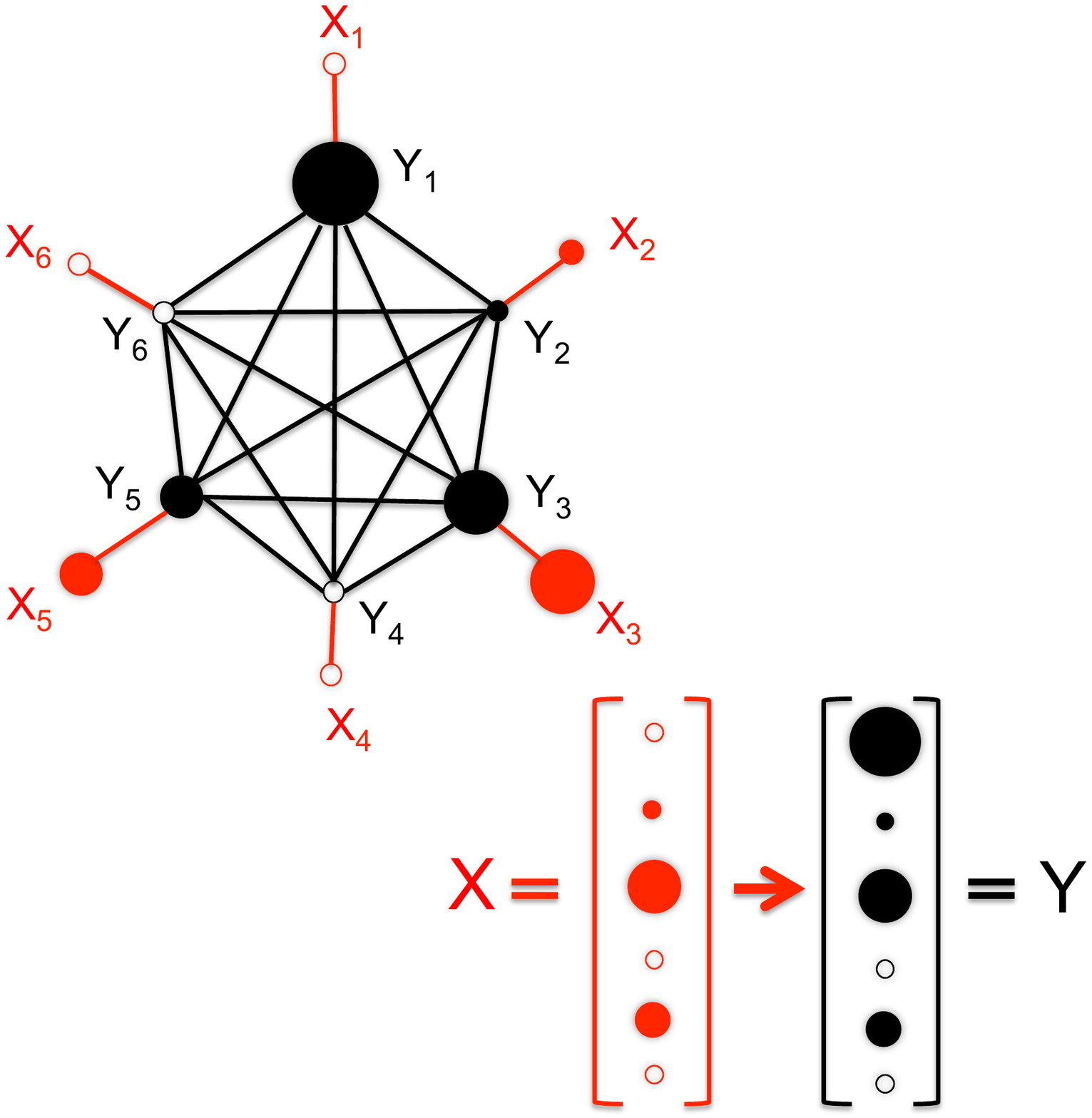}
 		\caption{The neural network representation of the  perfect  
	response described by the equation (\ref{evol2}) for $N=6$.  The red and black circles represent neurons of input and output layers respectively, with filled and empty circles denoting excited and unexcited neurons.  The size of a filled circle measures the  excitation level of a given neuron.  The largest black circle describes  the neuron $Y_1$ in the critically excited state. This enables the rest of 
$Y$-neurons to recognize an arbitrarily soft input $X$-stimulus pattern. } 		  		\label{recall}
 	\end{center}
 \end{figure}

  The state vector of the entire system evolves 
  as $\ket{t} = e^{-{i\over \hbar} \hat{H}t }\ket{\vec{Y}_{in}}\otimes\ket{\vec{X}_{in}}$.      
  The expectation value $Y_1$ does not change and 
   time-evolution of the expectation values $Y_{j\neq1}$
  (up to negligible corrections suppressed by powers of $g$)  
    takes the form
  \begin{equation} 
\bra{t} \hat{Y}_j \ket{t} \, = \,    Y_j(t) = X_j {\rm sin}^2\left ({tq \over 2\hbar}\right ) \, , 
  \label{evol1} 
   \end{equation}
or equivalently:
   \begin{equation} 
 \vec{Y}(t) \, = \, 
   \begin{pmatrix}
   {1\over g} \\
    0 \\
    ...\\   
    0  
\end{pmatrix} \, 
 +   {\rm sin}^2\left ({t q\over 2\hbar}  \right ) 
   \begin{pmatrix}
   0 \\
    X_2 \\
    ...\\   
    X_N  
\end{pmatrix} \,.  
\label{evol2}
\end{equation} 
 After time $t={\pi \hbar\over q}$,  the state of output $Y_{j\neq1}$-neurons  
   copies the input pattern $\vec{X}_{in}$.  So we are dealing with a perfect response (see Fig. \ref{recall}).  \\
   
   Thus, we see that  in the critical state  in which one of the neurons 
  is excited to $Y_1 = {1\over g}$, for the rest of the neurons
  $Y_{j\neq 1}$  the  pattern is retrieved under an arbitrarily soft external stimulus $X_j$.   The reason is that in the critical state -
  due to the excitatory synaptic connection with the excited neuron
  $Y_1$ - the rest of the neurons $Y_{j\neq 1}$ become effectively 
  {\it gapless} and can respond to an arbitrarily soft external stimulus. 
  This is analogous to the situation in which an ideal compass needle aligns itself to an arbitrarily weak external magnetic field.     \\
  
  Of course, the ability to bring the network of output neurons 
  in a desired state by an arbitrarily soft external stimulus, shows that  
 encoding of information is energetically cost-efficient. \\

   In sharp contrast,  under the influence of exact same input stimulus, 
 essentially no response takes place if the critical state is replaced 
 by a weakly-excited state. For example, let the initial state 
 be $\vec{Y}_{in} = \ket{0,0,0,...0}$. The evolution now takes the form  
   \begin{equation} 
 Y_j(t) = X_j \, { q^2 \over (1 + q^2)} {\rm sin}^2\left ({t \over 2\hbar} \sqrt{1 +  q^2} \right ) \, . 
  \label{evol2} 
   \end{equation}
  We see that the amplitude of the response is strongly suppressed by a factor $q^2$ and is rapidly oscillating.  
So, we are dealing with a very poor recognition.  

\section{Generic synaptic matrix}
 Generalization of the above phenomenon for non-universal synaptic matrix $W_{jk}$ and non-universal energy gaps $\epsilon_j$ in (\ref{H1}) is straightforward.  In order to see this, 
  let us arbitrarily split the set of $Y$-neurons in two subsets.  
 In order to distinguish them, we shall label 
 them by Latin and Greek indexes respectively: $\hat{Y}_{j}, \, j = 1,2, ...M$ and   $\hat{Y}_{\alpha}, \, \alpha = 1,2, ...N-M$, where 
 $M < N$ and is otherwise unspecified. \\
 
   Now assume that for a given 
 synaptic matrix there exists a split such that the set of following $M$ equations
 ($j = 1,2,...M$), 
  \begin{equation} 
    \epsilon_j \, - \, 2
   \sum_{\alpha =1}^{N-M}  W_{j\alpha} y_{\alpha} = 0  \, ,   
 \label{MEQ} 
 \end{equation}  
has a solution for some $y_{\alpha} = \xi_{\alpha}\,, 
\, \, (\alpha = 1,2,...N-M)$.   Then,  
the state 
\begin{equation} 
\ket{\vec{\xi}} \, = \, \ket{\xi_{1}, \xi_{2},...\xi_{N-M}, 0,0....0}
\label{MACRO}
\end{equation} 
 defines the neighbourhood of enhanced memory capacity. That is, there exist 
exponentially large number of states $\ket{\vec{Y}_{\xi}}$  -
obtained by excursions away from the state $\ket{\vec{\xi}}$ - that fit within an unit energy gap.  In such states the network can store an exponentially large number of patters. \\ 

 The  physical meaning of this phenomenon is clear. 
 In the state  (\ref{MACRO}) the thresholds for the excitations of the neurons
 $\hat{Y}_{j}, \, \,( j=1,2,...M)$ are zero, 
 \begin{equation} 
   \epsilon^{eff}_j = \epsilon_j \, - \, 2
   \sum_{\alpha =1}^{N-M}  W_{j\alpha} \xi_{\alpha} = 0  \, ,   
 \label{EEFF} 
 \end{equation}  
because $\xi_{\alpha}$-s solve the equations (\ref{MEQ}). 
 That is,   $\hat{Y}_{j}$-modes become effectively gapless. Thus, by varying their excitation levels, $y_j$,  we obtain  large number of  states at almost no energy cost. \\

  In order to count density of patterns,  let us restrict ourselves to the states  in which $y_{\alpha}$ are  kept frozen, $y_{\alpha} = \xi_{\alpha}$, whereas eigenvalues $y_i$ can take arbitrary  values. We denote such states as, 
 \begin{equation} 
\ket{\vec{Y}_{\xi}} \, = \, \ket{\xi_{1}, \xi_{2},...\xi_{N-M}, y_1,y_2,...y_M} . 
\label{MICRO}
\end{equation} 
The energy gap between the states 
$\ket{\vec{Y}_{\xi}}$ and $\ket{\vec{{\xi}}}$ is: 
 \begin{equation} 
    \Delta E =   
   \sum_{j, k =1}^{M}  W_{jk} y_jy_k  \,.    
 \label{HD} 
 \end{equation}
 Thus, the gap occupied by all possible states
 with $y_j\leqslant d$ is,  
  \begin{equation} 
    \Delta E = d^2  
   \sum_{j, k =1}^{M}  W_{jk}  \,,     
 \label{HD} 
 \end{equation}
 and the corresponding number of basic patterns is  
\begin{equation}  
 {\mathcal N}_{pattern}   = (d+1)^M \, .
 \label{NPM} 
 \end{equation} \\
 Thus, we again observe that by taking the synaptic weights 
 sufficiently weak, and $d$ and $M$ sufficiently large, we can 
 create a library of an arbitrarily large number of patterns, within 
 an arbitrarily small energy gap.  \\

For example, for    
$W_{jk} < {1\over M^2}$.  The number of patterns within $\Delta E  < 1$
is ${\mathcal N}_{pattern}   > 2^M$. 
 
 If synaptic connections 
 are taken to be weaker, the number of possible patterns increases. 
 For example, for $W_{jk} < {1\over M^4}$ any state 
 with $y_j\leqslant M$ fits  $ \Delta E  < 1$. The number of
 such patterns is
 ${\mathcal N}_{pattern}   > M^M$.  \\
 
 In general, for  
 $W_{jk} < {1\over M^2d^2}$ the number of patterns constrained by  
 $y_j\leqslant d$  that fit  
 within the unit energy gap is  ${\mathcal N}_{pattern}   > d^M$. \\

\begin{figure}
 	\begin{center}
        \includegraphics[width=0.4\textwidth]{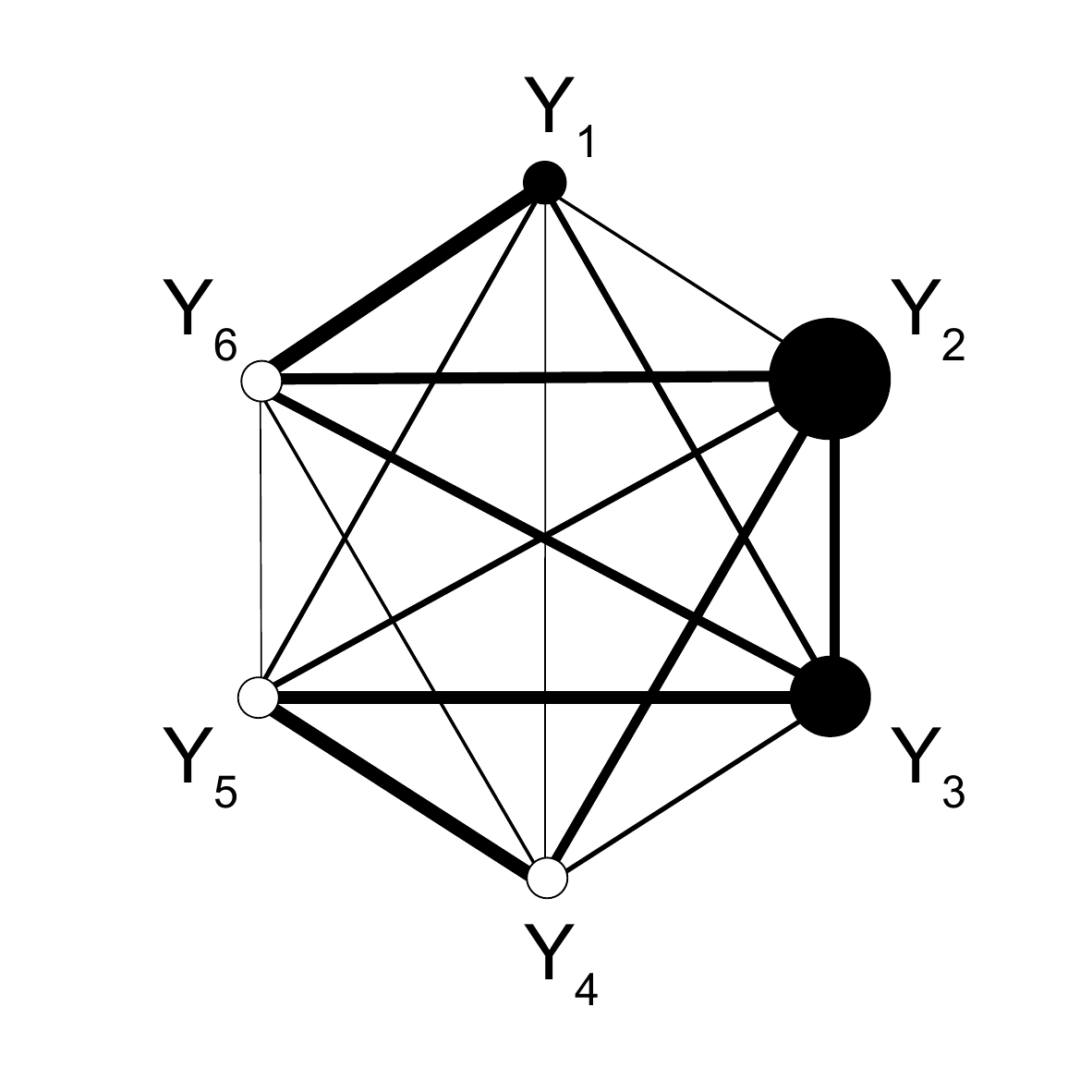}
 		\caption{The neural network representation of the
state $\vec{\xi} =
\begin{pmatrix}   
    10^{10}, & 3\cdot 10^{10}, & 2 \cdot  10^{10}, & 0,& 0, & 0    
\end{pmatrix} \,$ for the example given by the synaptic matrix (\ref{matrixG}) for $N=6$. The strength of each synaptic weights is proportional to the width of the corresponding line. The filled and empty circles denote excited and unexcited neurons.  The size of a filled circle measures the  excitation level of a given neuron. The neurons  $Y_1,Y_2$ and $Y_3$ are in critically excited states. 
This enables,  $Y_4,Y_5$ and $Y_6$ neurons to store a large number of patterns of the form  (\ref{YN6}) within a narrow energy gap.}
	  		\label{ExampleN6}
 	\end{center}
 \end{figure}

A concrete numerical example,  $N=6$ and $M=3$, 
 is described by Fig. \ref{ExampleN6}.
 The synaptic matrix and threshold energy vector are taken to be
    \begin{equation}
 \vec{W} = {10^{-10} \over 2}  
 \begin{pmatrix}
   0   & 2 & 4 &1&3 & 10\\
    2  &  0 & 6 & 6 &4& 7  \\
    4& 6& 0&  3 &  8 & 6&\\   
    1 & 6&3 &0& 9& 2& \\
     3&4& 8&  9 &  0 & 1&\\   
    10 & 7 &6 &2& 1& 0&    
\end{pmatrix},\, \,  
   \vec{\epsilon} =
 \begin{pmatrix}
    17 \\
     6\\
    11\\   
     25\\
      31\\
      43  
\end{pmatrix}.
\label{matrixG}
 \end{equation}
 A critical state is described by  the state vector \\
  
 $ \ket{\vec{\xi}} \, = \, \ket{10^{10}, 3\cdot 10^{10}, 2 \cdot  10^{10},0,0,0} $.\\
 
In this state, the neurons  $Y_{4}, Y_{5}, Y_{6}$ become effectively gapless. 
 Correspondingly,  within the unit energy gap, they can store  ${\mathcal N}_{pattern} \sim10^{15}$ distinct patterns  of the form, 
  \begin{equation} 
    \vec{Y} =
 \begin{pmatrix}   
    10^{10}\\   
     3\cdot 10^{10} \\
      2 \cdot  10^{10}\\
      y_4\\
      y_5\\
      y_6    
\end{pmatrix} \,,  \,\, \,   y_4,y_5,y_6 \lesssim 10^5 \,. 
\label{YN6}
 \end{equation}

 \section{Classical brains} 
 
  The question whether the quantum processing of information can play any significant role in the brain dynamics has been posed previously in different contexts (see, e.g., \cite{brainQ}).  For us, the most relevant is the formulation of this question in the context of quantum neural networks \cite{kak}. The critical quantum neural network discussed above may give some fresh perspective on this issue, due to the following reasons: \\
 
 \begin{itemize}
  \item  First, as we shall see, the critical enhancement of memory capacity fully persists in the classical limit of the network. Thus, this mechanism can be equally relevant for classical brains. 

  \item Secondly, it gives an example of how  - in otherwise-classical system - an intrinsically quantum way of information storage may become 
 operative. 
\end{itemize}

 We live in a quantum world. Yet, some processes are well-described classically. What is the measure of classicality? 
  Typically, a dynamics of the system is said to be {\it approximately classical} when the action of a process is much larger than $\hbar$. Such are the processes
  involving states with large occupation numbers of some quanta, i.e., 
  the highly excited states. \\
  
   For instance, a quantum oscillator 
 described by a number-operator $\hat{Y} = \hat{a}^{\dagger}\hat{a} $, 
 behaves almost classically when put in a highly excited  
 state $\ket{y}$, i.e, the state  with a large occupation number, $y \gg 1$.
 In fact, in such cases, the creation and annihilation operators 
 $\hat{a}^{\dagger},\hat{a}$
 can be replaced by 
 $c$-numbers, $\hat{a}^{\dagger} \rightarrow a^{*},  \hat{a}  \rightarrow a$, 
 where $|a| = \sqrt{y}$.  The error we commit  with such a replacement 
 is $\sim 1/y$.  This is the essence of the well-known Bogoliubov approximation \cite{bogoliubov}. \\
    
    Now, as we have seen, a large occupation number is an intrinsic property  of  the critical states discussed above. For example,  for the network 
 described by (\ref{H1}), the memory capacity becomes unbounded in the limit  $g \rightarrow 0$, in which the occupation number $y_1 = {1\over g}$ 
 becomes infinite.  Thus, the phenomenon of critical enhancement of memory capacity must have a well-defined classical limit. \\
 
  In order to show this, we can either use Bogoliubov approximation, or 
 use the basis of coherent states instead of the number eigenstates. 
  Since the coherent states are the eigenstates of the annihilation operators
 $\hat{a_j}$, we shall label them by the set of $\hat{a_j}$-eigenvalues,
 $\ket{\vec{Y}}_{coh} \equiv  \ket{\alpha_1,\alpha_2, ...\alpha_N}_{coh}$.
 That is, we have a relation,  $\hat{a_j}\ket{\vec{Y}}_{coh} = \alpha_j\ket{\vec{Y}}_{coh} $.  
 These eigenvalues  are the complex numbers $\alpha_j = \sqrt{Y_j} e^{i\theta_j}$, where $\theta_j$-s are the phases and  $Y_j$-s denote the expectation values of the corresponding number-operators over the state
 $\ket{\vec{Y}}_{coh}$.  That is, $|\alpha_j|^2 = Y_j$ measures the mean excitation level of $j$-th neuron 
 in this state. \\

In this basis, the critical state of the Hamiltonian  (\ref{H1})
can be represented by the coherent state with $\alpha_1 = 1/\sqrt{g}$ and $\alpha_{j\neq 1}=0$. We shall denote it by the following ket-vector, 
  $\ket{\vec{\xi}}_{coh} = \ket{1/\sqrt{g},0, ...0}_{coh}$. 
  Then, replacing the neuron $\hat{Y_1}$ in the Hamiltonian (\ref{H1}) by its expectation value, 
 $Y_1= {1/g}$, we get convinced that  
 in this state the neurons $\hat{Y}_{j\neq 1}$ are effectively gapless. 
 Thus, they can be used for storing patterns at low energy cost. \\
 
 To be more precise, the patterns can be stored in the set of coherent states, 
   $\ket{\vec{Y}}_{coh} = \ket{1/\sqrt{g},\alpha_2, ...\alpha_N}_{coh}$, which are obtained by small excursions, $|\alpha_{j\neq 1}| \ll  {1 \over \sqrt{g}}$, in the Fock space away from $\ket{\vec{\xi}}_{coh}$. Such states occupy the energy gap 
   analogous to (\ref{EEY}), 
    \begin{equation} 
    \Delta E = 
  {g \over 2} \sum_{1\neq j\neq k \neq 1}^{N} |\alpha_j\alpha_k|^2  \,.    
 \label{GAPAA} 
 \end{equation} 
 For small $g$, this gap can be made arbitrarily  narrow.  \\
     
  Thus, the information is now stored in a set of 
  $N-1$ complex numbers $(\alpha_2,...\alpha_N)$.  
   This is very similar to storing 
  patterns in the amplitudes and the phases of $N-1$ classical harmonic oscillators, each with almost zero frequency. \\
  
   Notice, $\alpha_j$ represent {\it continuous data}. 
 This may create a false impression that a range of available patterns is also continuous. 
  However, this is not the case. 
  Due to a well-known property of coherent states, only a discrete subset forms a  
  nearly-orthonormal set.   Indeed, the scalar product of two coherent states 
 $\ket{\vec{Y}}_{coh} = \ket{1/\sqrt{g},\alpha_2, ...\alpha_N}_{coh}$ and 
 $\ket{\vec{Y}'}_{coh} = \ket{1/\sqrt{g},\alpha_2', ...\alpha_N'}_{coh}$ is given by, 
   \begin{equation} 
   _{coh}\bra{\vec{Y}} \ket{\vec{Y'}}_{coh} = exp(-\sum_j|\alpha_j - \alpha_j'|^2) \,. 
  \label{orthog}
   \end{equation} 
Thus, the patterns encoded in orthogonal coherent states must satisfy, 
    \begin{equation} 
 \sum_j|\alpha_j - \alpha_j'|^2 \gg 1.
  \label{Distance}
   \end{equation} 
  Consequently, the number of distinct basic patterns is discrete.
  It is clear from 
 (\ref{GAPAA}) that the number of such patterns that fit within a fixed energy gap $\Delta E$ increases exponentially in the limit 
$g \rightarrow 0$.  Thus, in this limit  the network can store an unlimited amount of {\it classical information}. \\

  Thus,  the phenomenon of critical enhancement of memory   
  capacity carries over to a classical neural network. 
  Thus, a classical brain can make use of this phenomenon
  for creating an energetically cheap library of patterns, with a high precision 
  response to the soft external stimuli. What is required in the weak coupling 
  $g$, as the storage time scales as $1/g$. \\
  
  \section{Quantum entanglement} 
  
   The ``classicalized" critical brane network, in parallel to storing a large amount of classical data, has an ability to also store information in an intrinsically quantum way. In particular, the information is stored in entangled states. \\
     
   In order to explain this, it is convenient to think in terms of {\it macro} and {\it micro}  states. 
   Let us have a closer look at the degenerate states in which we can store a large density of patterns at the critical point. 
   For studying entanglement, it is more convenient to switch back to 
  the basis of  the number eigenstates,  labeled by $y_j$-s,  
 \begin{equation}
 \ket{{1 \over g}, y_2, ....y_N}\,.
  \label{library}
  \end{equation}  
  We are interested in the set of states that fits  within some narrow energy gap $\Delta E$.  For satisfying this condition - as it is clear from (\ref{EEY}) -   
  the occupation numbers $y_{j\neq 1}$ must be much smaller than 
 ${1 \over g}$. 
  All such states share one common {\it macroscopic}  characteristics: 
  A large occupation number (excitation level) of the first neuron,  $y_1 = {1\over g} \gg 1$. 
    On the other hand, the sets of the occupation numbers of the less-excited neurons,
  $(y_2,...y_N)$, represent the {\it microscopic}  characteristics of the states.  In this way, the states in (\ref{library}) can be considered 
  as the set of {\it micro-states}  that represent the same 
  {\it macro-state}. \\

   If the occupation numbers $y_{j\neq 1}$ are small,
  the information stored in such states in quantum.   
    For example, consider a library of patterns spanned  over 
 $2^{N-1}$ basic states with $y_{j\neq 1}=0,1$: 
 \begin{equation}
 \ket{{1 \over g}, 0,0,...,0}, \ket{{1 \over g}, 1,0,...,0},...,\ket{{1 \over g}, 1,1,...,1}\,. 
 \label{set}
 \end{equation}  
  The information can be stored in a highly entangled state formed by a superposition of the above basic vectors.   Here, we get an interesting scaling 
  relation.  As it is clear from (\ref{EEYdd}), such states 
  occupy the energy gap $\Delta E < gN^2$. Correspondingly,  the 
 time  over which the superposition stays intact 
 scales as, 
     \begin{equation} 
    \Delta t_{decoh} >  {1 \over gN^2} \,.       
 \label{DEC} 
 \end{equation} 
 Not surprisingly, the storage becomes permanent for $g \rightarrow 0$. 
  \\
 
    We  summarize with the following statements:   
 \begin{itemize}
  \item A classical brain that is away from criticality, can reliably store 
 information in the sets of states with large differences in occupation numbers.  
Moreover, both, the encoding and the retrieval of patterns are costly in energy, 
due to the absence of gapless modes.  Because of  the latter reason, the information stored in small occupation number differences is short-lived. 
Thus, such a brain can only store classical data.

  \item However, a brain put in the critical state, acquires  different abilities, due to emergence of gapless neurons.  
 Because of this, it can store information  in sets of  micro-states with large as well as small differences of the occupation numbers. Correspondingly, such a brain can store both,  classical and  quantum data, at a negligible energy cost.   
  
\end{itemize}

  The above features are independent 
 of our simple choice of number conserving Hamiltonian (\ref{H}).
 The ${1 \over g}$-scaling of the time  necessary for the change of the entanglement has been confirmed previously  \cite{mischa1} in critical models in which the interaction Hamiltonian  does not conserve individual occupation numbers.  So such a scaling appears to be a generic property of criticality.  
 As a matter of fact, it  has been shown in the earlier papers that the ground-state near the critical point is highly entangled \cite{daniel}, \cite{nico}, \cite{mischa1}.  \\
    
    Obviously, the information stored in highly entangled states cannot 
    be treated classically.  We are arriving to the following curious situation. \\

    On one hand, the brain network  described  by the Hamiltonian (\ref{H})   
    - when put in a  critical state - by all sensible measures  represents  a  {\it macroscopic} entity.  This is due to a 
  {\it macroscopically-large} occupation number of $Y_1$-mode in this state.
    Yet, in the very same macro-state,  the information can be encoded in a superposition  of the micro-states with small relative occupation numbers.  
  Moreover, the resulting state can be highly entangled.  The information carried by it, is {\it intrinsically quantum}. 
  Simultaneously, the storage time of such a quantum message is macroscopic, which is the behaviour characteristic of a classical information.  
    The property we are dealing with can be described as {\it macro-quantumness}.\\

     A necessary ingredient is a very weak synaptic connection
   not only among neurons, but also to the environment.  In such a case, 
   the thermal noise coming from the environment is not a problem, 
   since the thermalization time scales as $t_{thermal} \sim {1 \over g^2T}$
   (i.e., inverse interaction rate), which gives an extra factor ${1 \over g}$, as compared to the information storage time.  \\  
    
  Here, we shall very briefly comment  on the  
   issue of decoherence.  
 It is true that the decoherence time of superposition of 
 $2^{N-1}$ basic states can be macroscopically long due to the fact that 
 they are all crowded-up within a narrow energy gap. However, 
 this is again a trade-off. In order not to destroy 
 the gaplessness of modes, 
 we need a very weak coupling also to the environment. 
For example, for the individual biological neurons, such couplings are not small enough.  Implication of this fact, as Tegmark \cite{tegmark} showed,  is that the decoherence time of a neuron 
   is very short.  However, the $Y_j$-neurons in our network,
  may  correspond to some 
  collective excitations.  It is well-known that the coupling strength of the collective degrees of freedom can be much weaker, than the 
  couplings strength of the underlying ones.  The Goldstone bosons represent the  text-book example of this phenomenon. The 
  long-wavelength Goldstone modes can be arbitrarily weakly interacting.  \\     
  
 Because of this, the weakly interacting 
 particles, such as gravitons or Goldstone bosons, are suited the best 
 for forming artificial ``brains",  with critical memory enhancement.  
          The above indicates, why - if viewed as a neural network - a large black hole is  best suited for storing a quantum information. The universal gravitational coupling that plays the role of a synaptic connection, is extremely weak among gravitons of macroscopic wavelength.  For example, for a graviton of  $1$cm wavelength,  the analog of 
   synaptic strength would be $g \sim 10^{-66}$. \\

 \section{Brains and Black Holes}
 
  In this section we shall play with some numbers, which may be useful
  for comparing information storage efficiencies of black holes and brains.
 We must stress, however, that these numbers  
  {\it per se}, do not provide any evidence for similarities of the information storage mechanisms in these two systems.  \\

   It may sound surprising, but the energetic efficiency of information storage  by human brains is not that inferior to black holes.   A black hole of the 
   mass of  a human brain, approximately $ M_{BR} \sim 1$kg, would have a memory 
   storage capacity (measured by Bekenstein entropy \cite{Bek}), of order 
  ${M_{BR}^2 \over M_P^2} \sim 10^{16}$, where $M_P$ is the Planck mass, approximately $\sim 10^{28}$eV $\sim 10^{-8}$kg. This capacity is only by 
  $\sim 10^{5}$ times higher than a memory capacity of a human brain, estimated very roughly by counting the number of neurons in it ($\sim 10^{11}$). Notice however, that the elementary energy gap required 
for re-dialing the patterns stored in such a black hole, is 
$\epsilon_{BH} \sim 10^4$eV, which is much higher than the 
energy scale of the human neural activity. 
 Thus, although the human brain loses against an equal mass black hole in the competition 
  in memory capacity, it fully makes it up in the operational energy cost.

This comparison only takes into account the energy costs for redialing an average bit of information in the two systems. There is no other
well-defined sense in which we can compare them, since a black hole of 
a human brain mass is extremely short-lived, with the life-time of approximately $t_{BH} \sim 10^{-19}$sec.  
\\
       
       However, in what the black hole is absolutely superior to a human brain, is in the {\it compactness}  
       of the information storage.  A black hole of the mass of a human 
       brain, would have a size equal to
   $R_{BH} =  {M_{BR} \over M_P^2}  \sim 10^{-25}$cm, which is 
   approximately by a factor of $10^{-26}$ smaller than the human brain! \\
   
    According to the  idea of memory enhancement by criticality \cite{DG}, 
    this superiority is explained by the fact that the gravitational  coupling is
    universal and depends on an inverse square of the size of a compact system, $g = {\hbar^2 \over M_P^2 R_{BR}^2}$.  
   Because of this peculiar dependence, the synaptic weights  
   in a hypothetical ``brain" network in which the neurons interact only gravitationally,  secretly know about the size of the system.   
    So, such a brain reaches 
  criticality, when the size of the system becomes equal to its gravitational radius ($R_{BR} \sim M_{BR} / M_P^2$). This is the threshold beyond which 
  by laws of gravity the object becomes a black hole.  
   With a bit more elaboration (see, \cite{DG}), this line of reasoning  can provide a  {\it qualitative} explanation  - in the language of a 
  neural network -  why the  black holes saturate Bekenstein bound on information storage \cite{bekensteinBound}, which order of magnitude wise agrees with the bound  on information capacity by Bremermann \cite{Bremermann}. \\

 Of course, in the toy ``brain" models considered in this paper, the 
 strength of connections $g$ is just a parameter. What is crucial, however, 
  is that its inverse value sets the entropy of the critical state.  \\

  \section{Outlook}

    We have seen that in a simple quantum network with very weak excitatory  
   synaptic connections,  there exist the critically-excited states with large 
   memory capacity.  Moreover, the density of the available patterns 
  is an exponentially-growing function  of the  inverse strength of the synaptic weights.    
   This may sound counterintuitive, since naively one would expect that very weak synaptic connections should have no effect on the system's memory capacity.   However, this is a wrong intuition when we are dealing with 
  networks that are allowed to be in highly excited states.  For such networks, 
  the weak connections are the blessing, since the system can always find an excited state in which a set of neurons becomes effectively-gapless and free to store a large variety of patterns.  \\
  
  The entropy (i.e., the log of the number of micro-states) of such states is very high. 
  Already in the presented simple toy model,  
  it scales as the square-root of an inverse synaptic strength, 
  \begin{equation}
    Entropy \, \sim \, {1 \over \sqrt{g}} \,.
  \label{entropy} 
  \end{equation}

  In this paper, we did not address the question on what dynamics brings the system into the critical state. 
   This certainly can be caused by the external stimuli. 
  However, let us make the following remark.  Since, the critical state exhibits a maximal entropy, it is natural to expect that the network finds this state with a maximal probability, once enough energy is pumped in it. 
  Curiously enough, this argument is analogous to the one suggesting that 
  in very high energy particle collisions the black hole state is the most likely outcome, due to its maximal entropy. 
  \\

   Our point was to expose the existence of an energetically-cheap library 
   of patters with an exponentially large memory capacity and the ability of precise  response  to very soft input stimuli. 
 Our goal was not  to study particular algorithms for information
  processing.   
   Naturally, there is no obstacle to store information in highly entangled
   states from this library and implement algorithms for computing entanglement \cite{ental}. 
  In general, one can take the full advantage of quantum 
  superposition for neural network computing \cite{powerSuperpositioninNN}.   
  For example, algorithms for the quantum associative memory \cite{QuantAssociateMemory, QuantAssociateMemory1}, quantum pattern 
  recognition \cite{QuantumPatternRec}  and 
  quantum perception \cite{perception} can be implemented as 
 in the usual quantum neural networks, in which the existence of the 
 energy-cost-efficient qubits is assumed by design (for a review see, \cite{review}).
 The phenomenon we are after,  reveals one possible mechanism  for the brain  network to build such an energy-cost-efficient ``hardware".
  \\
 
 Another question that was beyond the focus of the present paper is learning. 
 In this respect, we can make the following remark.  The existence of a critical state
 in which the response to the input patterns is very precise, should be 
 helpful for the learning process. In such a state the system can train itself and 
 strengthen the needed set  of synaptic weights via 
 Hebbian learning \cite{Hebb}.  This can then facilitate the process of pattern-recognition also in the states that are away from criticality. 
   \\
   
    In conclusion, in the presence of weak negative synaptic weights in 
    a brain network, some highly excited states can acquire a large complexity and an ability   
  to store an exponentially large number of patterns.  This is similar to 
  how a weak attractive gravitational coupling produces states of high entropy and complexity in form of black holes.

\section*{Acknowledgements}

 We thank, Lasha Berezhiani, Allen Caldwell, Cesar Gomez and Tamara Mikeladze-Dvali, 
 for discussions.  
 Some of the presented models have been shared with 
 Dmitri Rusakov and Leonid Savchenko. It is a pleasure to acknowledge an ongoing communication with them.    
This work was supported in part by the Humboldt Foundation under Humboldt Professorship Award, ERC Advanced Grant 339169 "Selfcompletion", by TR 33 "The Dark Universe", and by the DFG cluster of excellence "Origin and Structure of the Universe". 

\appendix

\end{document}